\documentclass[twocolumn,prl,showpacs,preprintnumbers,amssymb]{revtex4-1}

\usepackage{epsfig}
\usepackage{dcolumn}
\usepackage{bm}

\usepackage{amsmath}
\usepackage{amssymb}
\usepackage{latexsym}
\usepackage{epsfig}
\usepackage{color}
\usepackage{mathtools}
\usepackage{youngtab}

\def\IC{\bf C}
\def\IZ{\bf Z}

\def\z2z2{$\IC^3/(\IZ_2\times\IZ_2)$}

\def\id{{\bf 1}}

\def\cp{\mbox{\bbbold C}\mbox{\bbbold P}}

\def\a{\alpha}

\def\b{\beta}

\def\d{\delta}\def\D{\Delta}

\def\k{\kappa}
\def\l{\lambda}

\def\p{\pi}

\def\s{\sigma}

\def\th{\theta}

\def\beq{\begin{equation}}\def\eeq{\end{equation}}
\def\beqa{\begin{eqnarray}}\def\eeqa{\end{eqnarray}}
\def\barr{\begin{array}}\def\earr{\end{array}}

\def\wt{\widetilde}

\def\ds {{\del \hspace{-6.4pt} \slash}\;}

 \let\br=\bigr

\def\bd{\begin{document}}
\def\ed{\end{document}}
\def\ba{\begin{array}}
\def\ea{\end{array}}
\def\bea{\begin{eqnarray}}
\def\eea{\end{eqnarray}}
\def\ft#1#2{{\textstyle{{\scriptstyle #1}\over {\scriptstyle #2}}}}
\def\fft#1#2{{#1 \over #2}}
\newcommand{\be}{\begin{equation}}
\newcommand{\ee}{\end{equation}}
\newcommand{\eq}[1]{(\ref{#1})}
\def\eqs#1#2{(\ref{#1}-\ref{#2})}
\def\det{{\rm det\,}}
\def\tr{{\rm tr}}
\newcommand{\ho}[1]{$\, ^{#1}$}
\newcommand{\hoch}[1]{$\, ^{#1}$}
\def\ra{\rightarrow}

\def\Xh{\hat{X}}
\def\ah{\hat{a}}
\def\xh{\hat{x}}
\def\yh{\hat{y}}
\def\ph{\hat{p}}
\def\G{{\cal G}}
\def\Dth{{\Delta_\th}}

\def\bk{{\bf k}}
\def\bx{{\bf x}}
\def\br{{\bf r}}
\def\tr{{\rm tr \,}}
\def\Tr{{\rm Tr \,}}
\def\diag{{\rm diag \,}}
\def\tg{{\rm tg \,}}
\def\NPB#1#2#3{Nucl. Phys. B {\bf #1} (19#2) #3}
\def\PLB#1#2#3{Phys. Lett. B {\bf #1} (19#2) #3}
\def\PLBold#1#2#3{Phys. Lett. {#1B} (19#2) #3}
\def\PRD#1#2#3{Phys. Rev. D {\bf #1} (19#2) #3}
\def\PRL#1#2#3{Phys. Rev. Lett. {\bf #1} (19#2) #3}
\def\PRT#1#2#3{Phys. Rep. {\bf #1} C (19#2) #3}
\def\MODA#1#2#3{Mod. Phys. Lett.  {\bf #1} (19#2) #3}
\def\ov{\overline}

\def\preal{{\rm Re\,}}
\def\pim{{\rm Im\,}}

\def\ds{\displaystyle}
\def\yzero{\smash{\hbox{$y\kern-4pt\raise1pt\hbox{${}^\circ$}$}}}
\def\p{\partial}
\def\a{\alpha}
\def\b{\beta}
\def\g{\gamma}
\def\d{\delta}
\def\beq{\begin{equation}}
\def\eeq{\end{equation}}
\def\beqa{\begin{eqnarray}}
\def\eeqa{\end{eqnarray}}
\def\Om{\Omega}
\def\om{\omega}
\def\th{\theta}
\def\vt{\vartheta}
\def\vphi{\varphi}
\def\-{\hphantom{-}}
\def\ov{\overline}
\def\s2{\frac{1}{\sqrt2}}
\def\wh{\widehat}
\def\wt{\widetilde}
\def\oh{\frac{1}{2}}
\def\tr{{\rm tr \,}}
\def\Tr{{\rm Tr \,}}
\def\diag{{\rm diag \,}}
\def\vac{|0 \rangle}
\def\vm{\relax{n_{\text{v}}}}
\def\cc{{\cal C}}
\def\ck{{\cal K}}
\def\ci{{\cal I}}
\def\cu{{\cal U}}
\def\cg{{\cal G}}
\def\cn{{\cal N}}
\def\cam{{\cal M}}
\def\cp{{\cal P}}
\def\ct{{\cal T}}
\def\cv{{\cal V}}
\def\cz{{\cal Z}}
\def\ch{{\cal H}}
\def\cf{{\cal F}}
\def\tv{\tilde v}
\def\Dsl{\,\raise.15ex\hbox{/}\mkern-13.5mu D} 
\def\IZ{Z\kern-.4em  Z}
\def\id{{\rm 1}}

\def\ti{\times}
\def\til{\tilde}
\def\eps{\epsilon}
\def\k{\kappa}
\def\A{\Arrowvert}
\def\cw{{\cal W}}
\def\G{\Gamma}
\def\car{{\cal R}}
\def\l{\lambda}
\def\raw{\rightarrow}
\def\Raw{\Rightarrow}
\def\inte{{\bf Z}}
\def\cpx{{\bf C}}
\def\real{{\bf R}}
\def\Lam{\Lambda}
\def\D{\Delta}
\def\cb{{\cal B}}
\def\ca{{\cal A}}

\begin{document}

\preprint{MAD-TH-13-05~
MPP-2014-5}

\title{Probing Hidden Sectors with St\"uckelberg $U(1)$ gauge fields}

\author{Wan-Zhe Feng$^{1,2}$, Gary Shiu$^{1,3}$, Pablo Soler$^{1,3}$, and Fang Ye$^{1,3}$}
\affiliation{\small\slshape  $^{1}$ Center for Fundamental Physics and Institute for Advanced Study, Hong Kong University of Science and Technology, Hong Kong \\
     $^{2}$ Max--Planck--Institut f\"ur Physik (Werner--Heisenberg--Institut),
80805 M\"unchen, Germany\\
$^{3}$ Department of Physics, 1150 University Avenue, University of Wisconsin, Madison, WI 53706, USA}
\begin{abstract}
We propose a framework in which visible matter interacts with matter from a hidden sector through mass mixings of Stueckelberg $U(1)$ gauge fields.
In contrast to other $Z'$ mediation scenarios, our setup has the added appealing features 
that  (i) the choice of $Z'$s can be significantly broadened without necessarily introducing unwanted exotic matter and (ii) there can be sizable tree-level interactions between the visible and hidden sectors. 
String theory embeddings of this scenario and their phenomenological features are briefly discussed.
\end{abstract}
\maketitle

\section{Introduction}

The observational evidence for dark matter (DM) 
is perhaps currently the most compelling case for  physics beyond the Standard Model.
Other than its gravitational influence, how this dark sector interacts with ordinary matter remains a complete mystery.
Thus, determining the channels through which the visible and dark sectors communicate with each other not only helps uncover new forces and symmetries in Nature, it
has deep and direct impact on 
the experimental program of DM searches.

One simple way to realize the DM sector is broadly known as 
the hidden sector scenario. In its minimal form, it consists 
 of a visible sector with the (Minimal Supersymmetric) Standard Model ((MS)SM) matter content $\Psi_{\text{v}}$ and gauge group $SU(3)_c\times SU(2)_L\times U(1)_Y$; and a hidden sector with gauge group $G_{\text{h}}$ and matter content $\Psi_{\text{h}}$ charged under it, but neutral under the visible group:
\begin{eqnarray}\label{higgsportal}
\text{Group}&~~~~~&SU(3)_{c}\times SU(2)_{L}\times U(1)_Y~~\times~~ G_{\text{h}}\\[-10pt]
&~~~~~&\underbrace{ \hphantom{SU(3)_{c}\times SU(2)_{L}\times U(1)_Y}}_{\Psi_{\text{v}}} \hphantom{~~\times\,\,\,}\underbrace{ \hphantom{G_{\text{h}}}}_{\Psi_{\text{h}}}\nonumber\\[-16pt]\nonumber \vphantom{\underbrace{\text{Group}}}_{\text{Matter}}
\end{eqnarray}
In this setup, the Higgs boson plays a  special role. 
The only super-renormalizable coupling of the SM is its mass term $\mu^2 H^{\dagger}_{\text{v}} H_{\text{v}}$, which hence admits renormalizable couplings to hidden sectors scalars $\lambda \Phi_{\text{h}}^{\dagger}\Phi_{\text{h}} H_{\text{v}}^{\dagger} H_{\text{v}}$. Thus, the Higgs boson is a simple portal into hidden sectors \cite{Patt:2006fw}
(several other portals have been proposed, see e.g.~\cite{Essig:2013lka}).

In this work 
we point out that in the presence of heavy $Z'$ bosons, there is yet another efficient portal, in the sense that the interactions between the visible and dark sectors
appear also at the renormalizable level.
Given that heavy $Z'$ bosons appear generically in beyond the SM physics \cite{Langacker:2008yv}, as well as string constructions \cite{Stringy-Z'}, we expect our scenario to have wide applicability. Our setup amounts to extending the structure of~\eqref{higgsportal} by extra $U(1)$ factors both in the visible and in the hidden sectors:
\begin{eqnarray}\label{stuckportal}
\!\!\!SU(3)_{c}\!\times\! SU(2)_{L}\!\times\! U(1)_Y\!\!\times\! U(1)_{\text{v}}^n~\times~U(1)_{\text{h}}^m\!\times\! \tilde{G}_{\text{h}} \\[-10pt]
\!\!\!\underbrace{ \hphantom{SU(3)_{c}\!\times\! SU(2)_{L}\!\times\! U(1)_Y\!\!\times\! U(1)_{\text{v}}^n}}_{\Psi_{\text{v}}} \hphantom{~\times~}\underbrace{ \hphantom{U(1)_{\text{h}}^m\!\times\!\tilde{G}_{\text{h}}}}_{\Psi_{\text{h}}}\nonumber
\end{eqnarray}
Here, $\tilde{G}_{\text{h}}$ represents the semi-simple part of the hidden gauge group. $U(1)_{\text{v}}^n$ are $n$ abelian gauge groups to which the (MS)SM matter fields couple, and whose gauge bosons are massive. $U(1)_{\text{h}}^m$ are $m$ abelian gauge factors (some of which could be massless) to which only hidden matter couples.

We will argue that
$Z'$ bosons are natural portals 
between
the visible and 
hidden sectors. More concretely, the mass matrix for the gauge bosons $(A_{\text{v}}^n~A_{\text{h}}^m)$ can be non-diagonal, and upon diagonalization, may yield `physical' $Z'$ eigenstates (those with diagonal kinetic and mass terms) that couple simultaneously to both matter sectors. This mass mixing is a tree-level effect that can be the dominant interaction between separated sectors, provided the 
associated
$Z'$ bosons are light enough~\cite{kinetic}.

The scenarios 
discussed here find a natural implementation and motivation in D-brane models, 
where massive $U(1)$ bosons are ubiquitous (for reviews see e.g. \cite{Ibanez:2012zz,Blumenhagen:2005mu,Blumenhagen:2006ci,Marchesano:2007de}). 
However, the mechanisms we describe can be employed in a more general context,
and so we begin with a low energy description of the setup.
We  discuss briefly the corresponding string theory ingredients in the last section. More details are given in a companion paper~\cite{us}, where the first global embeddings of this genuine hidden sector scenario (with no exotics) into string theory are described.

\section{$\bf{U(1)}$ masses and their mixing}
Let us first review the origin of $U(1)$ gauge boson masses. The mechanism involves pseudo-scalar periodic fields $\phi^i$ (normalized so that $\phi^i\equiv\phi^i+1$) that transform non-linearly $\phi^i\to\phi^i+k^i_a\Lambda_a$ under gauge transformations $A_a\to A_a+d\Lambda_a$. Their kinetic terms read
\begin{equation}\label{stuckelberg}
{\cal{L}}=-\frac{1}{2}\, G_{ij}\left(\partial\phi^i-k^i_aA_a\right)\big(\partial\phi^j-k^j_bA_b\big)\,,
\end{equation}
where $G_{ij}$ corresponds to the (positive definite) metric on the space of fields $\phi^i$. In our normalization it has mass dimension 2. One can directly read off the mass matrix that the gauge bosons acquire by absorbing these axions:
\begin{equation}\label{masses}
(M^2)_{ab}=G_{ij}k_a^ik_b^j=\left(K^{\,T}\cdot G\cdot K\right)_{ab}\,.
\end{equation}

We notice here the important fact that one can always use a normalization of the gauge fields such that all the entries of the $K$ matrix as well as all the $U(1)$ charges of matter fields in the system are integers. Such quantization of charges, together with the periodicity of axions, simply describes the fact that the gauge groups are compact ($U(1)$ rather than $\mathbb{R}$), which is a requirement  imposed by quantum gravity on any effective gauge theory to which it can be coupled~\cite{Banks:2010zn,Shiu:2013wxa}. Henceforth, we assume that such a normalization has been taken, and refer to the integers $k^i_a$ as the $U(1)$ charges of the axions.

Both the St\"uckelberg and the Higgs mechanisms can be described by~\eqref{stuckelberg}. We focus here on the former case because of its connection to 
the Green-Schwarz (GS) anomaly cancellation mechanism, and because of its prominent appearance in string theory models.
As we shall see, the St\"uckelberg mechanism also offers more options for the extra $U(1)$'s, without the need of introducing unwanted matter exotics.

We already see from~\eqref{masses} that non-diagonal mass terms can easily arise and connect visible and hidden $U(1)$'s, both through a non-diagonal metric $G$ and also by having axions simultaneously charged under both sectors. In particular, such mixed charges, being integral, can generate highly non-diagonal mass matrices, and hence lead to a strong mixing of $U(1)$ bosons. This is a very appealing (and as we will see well motivated) mechanism to connect the Standard Model with hidden sectors that we call the ``St\"uckelberg portal''.

\vspace{8pt}

{\bf{A simple model:}}
Let us illustrate this portal with a simple example. We consider an extension of the SM by an extra $U(1)_{\text{v}}$, and a hidden sector with an abelian group $U(1)_{\text{h}}$. We arrange their gauge bosons in a vector $(A_{\text{v}}\,A_{\text{h}})$, and consider two axions that transform under both groups with a generic matrix of charges
\begin{equation}\label{charges}
K=\left(
\begin{array}{cc}
a & b \\
c  & d
\end{array}\right)~~~~a,b,c,d\in{\mathbb Z}~~~~\det(K)\neq 0
\end{equation}
We assume for simplicity that the axion metric is diagonal with two mass scales $M$ and $m$ associated to the axions, that is, we take $G={\text{diag }}(M^2, \,m^2)$.

As we mentioned before~\cite{kinetic}, we work in the approximation that the kinetic term is diagonal ${\cal{L}}_{\text{kin}}\sim g_a^{-2} F_a^2$. It can be written canonically by reabsorbing the coupling constants into the gauge fields $A_a\to g_a A_a$. The final gauge boson mass matrix then reads
\begin{eqnarray}\label{example1}
M^2_{U(1)}&=&\left(
\begin{array}{cc}
g_{\text{v}} & 0 \\
0 & g_{\text{h}}
\end{array}\right)
K^{\,T}\!\!\cdot \!G\!\cdot\! K
\left(
\begin{array}{cc}
g_{\text{v}} & 0 \\
0 & g_{\text{h}}
\end{array}\right) \nonumber\\&=&
\left(
\begin{array}{cc}
g_{\text{v}}^2 (a^2 M^2+c^2 m^2) & g_{\text{v}} g_{\text{h}} (a \,b\, M^2 + c\,d\, m^2)\\
g_{\text{v}} g_{\text{h}} (a\, b\, M^2 + c\,d\, m^2) & g_{\text{h}}^2 (b^2 M^2 + d^2 m^2)
\end{array}\right)\nonumber
\end{eqnarray}
It is clear that the eigenvectors of this highly non-diagonal matrix, i.e. the physical $Z'$ bosons, will be linear combinations of $A_{\text v}$ and $A_{\text h}$. Hence they will couple to matter currents from both the visible and the hidden sectors, and generically they will do so with similar strengths. To further illustrate this point let us take the limit $\epsilon\equiv m/M\ll 1$, in which expressions simplify. The physical bosons are expressed up to order ${\cal O}(\epsilon^2)$ as   
\begin{eqnarray}\label{eigenstates}
Z'_m&\approx& g_m\left(b \, \frac{A_{\text{v}}}{g_{\text{v}}^2} -  a\, \frac{A_{\text{h}}}{g_{\text{h}}^2}\right) ~~~~{\text{Mass}}(Z'_m)\approx m\, g_m\, {\text{det}}(K)\nonumber\\
Z'_M&\approx& g_M\left( a\, \frac{A_{\text{v}}}{g_{\text{v}}^2} +  b\,\chi\frac{A_{\text{h}}}{g_{\text{h}}^2} \right)~~~~{\text{Mass}}(Z'_M)\approx M\frac{g_{\text{v}}^2}{g_M}
\end{eqnarray}
where we have defined the couplings
\begin{equation}
\frac{1}{g^2_m}\equiv\frac{b^2}{g_{\text{v}}^2}+\frac{a^2}{g_{\text{h}}^2},\qquad \frac{1}{g^2_M}\equiv\frac{a^2}{g_{\text{v}}^2}+\chi^2\frac{b^2}{g_{\text{v}}^2},\qquad \chi\equiv \frac{g_{\text{h}}}{g_{\text{v}}}=\frac{g_m}{g_M}\nonumber
\end{equation} 
Their interactions with visible and hidden matter currents $J_{\text{v}}$ and $J_{\text{h}}$, are written as
\begin{eqnarray}
{\cal{L}}_{\text{int}}&=&g_{\text{v}} A_{\text{v}} J_{\text{v}} +g_{\text{h}} A_{\text{h}} J_{\text{h}} \nonumber\\
&\approx&g_m Z'_m (b\,J_{\text{v}} - a\,J_{\text{h}})+g_M Z'_M(a\,J_{\text{v}}+\chi^2 b \,J_{\text{h}})\,,\nonumber
\end{eqnarray}
again, up to order ${\cal O}(\epsilon^2)$ corrections. We can see that both the hidden and the visible sector couple with similar strength to the lightest gauge boson, and to the heavier one as well (provided $\chi\sim{\cal{O}}(1)$). We consider this class of models to be phenomenological very interesting, they can indeed  be tested at the LHC if the masses of the lightest $Z'$ lies around the TeV scale. 

We stress here that the large mixing between hidden and visible bosons of~\eqref{eigenstates} is a consequence only of the mixed axionic charges in~\eqref{charges} and does not rely at all on the approximation $m/M\ll 1$ that was assumed only to simplify the resulting expressions. In the case $m\approx M$, the results are slightly more complicated, but both physical $Z'$ gauge bosons are still (generically) largely mixed combinations of $A_{\text{v}}$ and $A_{\text{h}}$. 

\vspace{8pt}

\begin{figure}[!htp]
\centering
\includegraphics[width=240pt]{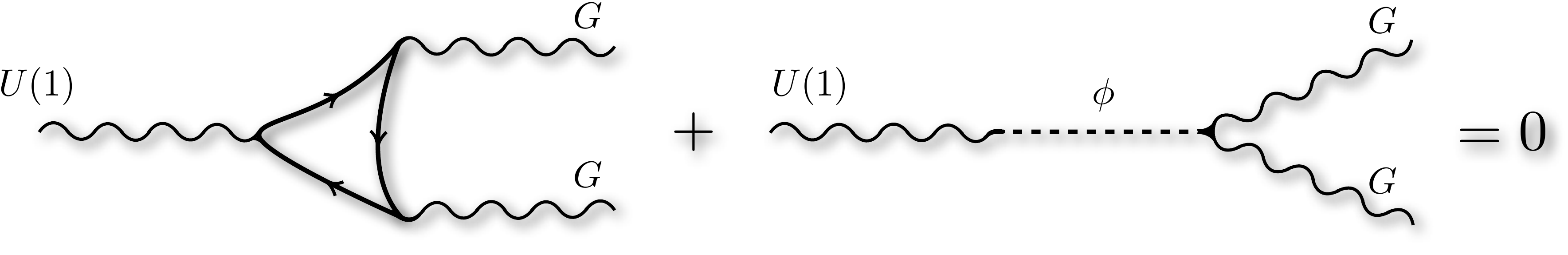}
\caption{\footnotesize GS mechanism for $U(1)-G^2$ anomaly cancellation.}\label{GS}
\end{figure}

{\bf{Anomalies:}}
A first reason to focus on the St\"uckelberg, rather than on the Higgs mechanism is the tight connection between St\"uckelberg axions and the GS mechanism for anomaly cancellation. Basically, any gauge triangle anomaly involving a massive $U(1)$ can be cancelled by contributions from the St\"uckelberg axions if these couple appropriately to the gauge bosons (fig.~\ref{GS}). The first vertex in the axionic diagram comes from the lagrangian~\eqref{stuckelberg}, which is present for any massive $U(1)$, while the second vertex corresponds to a coupling of the form $\phi\, F_G\wedge F_G$.

We will assume here that the GS-mechanism is in force, and hence we will not be restricted to take $U(1)_{\text{v}}$ to be $B-L$ or anomaly-free family-dependent symmetries, nor do we need to consider exotic matter with SM couplings to cancel anomalies~\cite{Anomaly}.

\section{Comments on the phenomenology}

From a phenomenological perspective, this class of models consists of an extension of 
the (MS)SM by extra massive $U(1)$ groups, i.e. $Z'$ bosons, to which both visible and hidden matter couple. The phenomenology of heavy $Z'$ has been vastly studied~\cite{Langacker:2008yv}. In this paper, however, we note that such bosons can couple naturally  to hidden sectors and represent a very well motivated portal into them (see \cite{Alves:2013tqa,Arcadi:2013qia} for a recent phenomenological discussion of similar models). 

The phenomenology of our scenario depends drastically on the particular visible $U(1)_{\text{v}}$ under consideration. Given the fact that the GS mechanism is in force cancelling anomalies, there are many possibilities, such as $U(1)_B,~U(1)_L,~U(1)_{PQ}$, lepto- or quarko-phobic, etc~\cite{anomalous}. 

Other factors upon which the models depend are the couplings $g_{\text{v}}$ and $g_{\text{h}}$, the masses of dark particles coupled to $U(1)_{\text{h}}$, and especially the final mass of the lightest physical $Z'$. In fact, since the latter will generically couple with significant strengths to matter from the visible sector, the rather stringent LEP and LHC constraints require considering $Z'$ bosons whose masses are at least around the TeV scale. Obviously, the most interesting scenario would include a $Z'$ whose mass is within the reach of LHC. 

In setups with extended abelian sectors, lower bounds on $Z'$ masses also arise from EW constraints on $Z-Z'$ mixing. Although in the field theory models we have discussed so far such mixings need not be considered, these arise generically in string theory implementations, as discussed in~\cite{Ghilencea:2002da,us}.

Additionally, in our setups, matter fields from the hidden sector are natural DM candidates. They could annihilate through the $Z'$ poles to produce pairs of SM fermions: $\overline{\psi}_{\text h}+\psi_{\text h}\to Z' \to \overline{\psi}_{\text v}+\psi_{\text v }$. Through this process, and for an appropriate range of masses and couplings, one can reduce the density of primordial hidden particles and satisfy the current DM relic density. This interesting possibility will be explored in depth in~\cite{us}.

\vspace{10pt}

One particular characteristic of $Z'$ bosons arising from mass mixing is that their couplings to particles from different sectors are generically not quantized with respect to each other. This means that gauge invariant operators cannot be constructed out of gauge variant components from different sectors~\cite{invariants}. 
This is in contrast with the generic situation one would encounter if the SM was extended by a single extra $U(1)$ to which the hidden sector would couple directly (and hence with quantized charges). 

One application of our setup is the possible mediation of supersymmetry breaking from the hidden sector (where supersymmetry breaking could be triggered, e.g. by strong dynamics from the semi-simple part $\tilde{G_{\text{h}}}$ of the hidden group) to the MSSM by $Z'$ bosons. Such a mechanism was proposed in~\cite{Langacker:2007ac,Langacker:2008ip}. 
Our scenario is nonetheless different in several respects. The cancellation of $U(1)$ anomalies by the GS mechanism allows us to construct models where the mediation is purely through the $U(1)$ bosons without the need of introducing matter exotics. Moreover, the strong mixings between the visible and hidden sectors can lead to more pronounced signatures.

In a certain sense, our setup is similar to that considered in~\cite{Verlinde:2007qk}, since both involve mass mixing of $U(1)$ bosons coupled to axions. However,  
\cite{Verlinde:2007qk} involves massless gauge bosons (see also~\cite{Nath,Kors:2004dx,Cheung:2007ut,Feldman:2007wj}), and hence has very different features. In particular, matter from the hidden sector could easily carry exotic couplings to the SM.

For an appropriate confining hidden sector, our models can be viewed as a `hidden valley' scenario~\cite{Strassler:2006im,Strassler:2006qa,Strassler:2006ri}, where the barrier energy scale is set by the mass of the lightest $Z'$. The phenomenology of a particular simple case with $U(1)_{\text{v}}$ taken as a (anomaly free) linear combination of B-L and hypercharge was explored in detail in~\cite{Han:2007ae}.
Here we note that mass mixing naturally results in such models, and furthermore the choice of $U(1)$'s in such scenarios can be significantly broadened.

We finally mention here the possibility to obtain a small mass mixing between $U(1)$ bosons from different sectors by an almost diagonal metric $G$ (e.g. with non-diagonal terms induced by loop-effects)~\cite{us}. This small mixing is in contrast to the large mixing induced by mixed axionic charges (which is generically large due to their integrality) and can have interesting applications to `hidden photon' scenarios where the mass of the hidden $U(1)$ is very small ($m_h\lesssim{\text{GeV}}$) and large mixings are ruled out by experiments~\cite{Jaeckel:2010ni,Essig:2013lka}.

\section{String theory implementation}
In this final section we describe the scenario presented above in terms of D-brane models of type II string theory. As we will see, the St\"uckelberg portal finds a natural implementation in these setups. We focus in particular on models of intersecting D6-branes in type IIA, where the geometrical intuition in terms of homology cycles of the compactification space is clearer (see e.g.~\cite{Ibanez:2012zz,Blumenhagen:2005mu, Blumenhagen:2006ci, Marchesano:2007de}. 

Given a four dimensional type IIA orientifold compactification, gauge theories arise from stacks of D6-branes that wrap three-cycles of the internal manifold $\bf{X_6}$, 
and span the four non-compact Minkowskian directions.
The 4d gauge group living on a generic stack of $N$ coincident D6-branes is (locally) $U(N)\cong SU(N)\times U(1)$, and contains an abelian $U(1)$ factor.
The ubiquitous presence of such groups gives a strong motivation to consider $U(1)$ extensions of the (MS)SM.

Let us take a basis of three-cycles $\{[\alpha^i],[\beta_i]\}_{i=0,\ldots,h_{2,1}}$ of ${\bf{X_6}}$, with $[\alpha^i]$ even and $[\beta_i]$ odd under the the orientifold projection, whose non-zero topological intersection numbers read
\begin{equation}
[\alpha^i]\cdot[\beta_j]=-[\beta_j]\cdot[\alpha^i]=\delta_j^i\,.
\end{equation}
One can express the three-cycles wrapped by a given stack of $N_a$ branes in terms of this basis as
\begin{equation}\label{wrappings}
[\Pi_a] = s_{ai}[\alpha^i]+r_a^{\,j}[\beta_j]  \,.
\end{equation}
Matter fields arise at the intersection of two such stacks, where 
there are chiral fermions that transform under the bifundamental representation $(\,\tiny\yng(1)\,,\overline{\tiny{\yng(1)}}\,)_{(+1,-1)}$ of the gauge group $SU(N_a)\times SU(N_b) \times U(1)_a\times U(1)_b$. Their multiplicity is given by the intersection numbers
\begin{equation}
I_{ab}\equiv[\Pi_a]\cdot[\Pi_b]=s_{ai}\,r_b^i- r_a^{\,i}\,s_{bi}\,.
\end{equation}

The axions $\phi^i$ we will be discussing come from the reduction of the 10d Ramond-Ramond (RR) three-form $C_3$ along orientifold-even three-cycles $[\alpha^i]$ of the compactification space:
\begin{equation}
\phi^i\equiv \int_{[\alpha^i]}C_3~,~~~~~~i=0,\ldots h_{2,1}({\bf X}_6)\,.
\end{equation} 
It can be seen from the reduction of the 10d Chern-Simons action that these fields have shift transformations under the abelian factors $U(1)_a\subset U(N_a)$: 
\begin{equation}
A^a\to A^a+d\Lambda^a~~\qquad \phi^i\to \phi^i+N_a \,r_a^{\,i}\Lambda^a\,.
\end{equation}
Hence, we can identify the axionic charges as $k_a^i\equiv N_a r_a^i$ (notice that $k_a^i\in{\mathbb{Z}}$). The kinetic term for these axions ${\cal{L}}_{\text{kin}}\sim G_{ij} D\phi^i D\phi^j$ contains a St\"uckelberg coupling like~\eqref{stuckelberg}, where the metric $G$ is the complex structure (c.s.) moduli space metric of the internal space ${\bf X_6}$.

\vspace{8pt}

The crucial point in realizing the St\"uckelberg portal is that, while charged chiral fermions only appear at brane intersections, RR axions come from closed strings that propagate in the bulk and can hence interact with different sectors, even if they are geometrically separated in the internal space. One can easily construct setups in which the visible and hidden sectors arise from stacks that do not mutually intersect, so a structure like~\eqref{stuckportal} is reproduced. By appropriately choosing the wrapping numbers $s_{ai}$ and $r_a^i$, one can obtain RR axions $\phi^i$ that are charged simultaneously under $U(1)$'s from different sectors and generate accordingly non-diagonal mass matrices for the corresponding gauge bosons:
\begin{eqnarray}
& \hphantom{SU(3)_{c}\times SU(2)_{L}\times U(1)_Y \times U} \overbracket[0.5pt][7pt]{ \hphantom{(1)_{2v}~~\times~~U}}^{\phi^i} \hphantom{(1)_h\times G_h}\nonumber\\[-5pt]
&SU(3)_{c}\times SU(2)_{L}\times U(1)_Y \times U(1)^n_{\text{v}}~~\times~~ U(1)^m_{\text{h}}\times G_{\text{h}}\nonumber\\[-10pt]
&\underbrace{ \hphantom{SU(3)_{c}\times SU(2)_{L}\times U(1)_Y \times U(1)^n_{v}}}_{\Psi_{\text{v}}} \hphantom{~~\times\,\,\,}\underbrace{ \hphantom{U(1)_h\times G_h}}_{\Psi_{\text{h}}}\nonumber
\end{eqnarray}

\vspace{8pt}

Let us illustrate this $U(1)$ mixing mechanism with a toy model that reproduces the example presented in eq.~\eqref{charges}. We take an inner space with $h_{2,1}({\bf{X_6}})\geq 3$ (e.g. a six-torus $\bf{T^6}$), so that there exist at least three even $[\alpha^i]$ and three odd $[\beta_i]$ three-cycles. We consider two branes, a `visible' and a `hidden' one that host a gauge group $U(1)_{\text v}\times U(1)_{\text h}$, wrapping the following three-cycles:
\begin{eqnarray}\label{toywrappings}
\begin{array}{c}
[\Pi_{\text{v}}]= [\alpha^0]+a[\beta_2]+c[\beta_3]\\ 
[5pt]\phantom{}[\Pi_{\text{h}}]= [\alpha^1]+b[\beta_2]+d[\beta_3]\,.
\end{array}
\end{eqnarray}
It is straightforward to see that both branes do not intersect, i.e. $[\Pi_{\text v}]\cdot[\Pi_{\text h}]=0$ (nor do their orientifold images). Hence there is no chiral matter charged simultaneously under both $U(1)$ factors. In order to include chiral matter in the system, we could add a stack of branes along $[\beta_0]$ which would intersect $[\Pi]_{\text v}$ but not $[\Pi]_{\text h}$, yielding `visible' matter; and/or a stack along $[\beta_1]$ leading equivalently to charged matter on the `hidden sector', uncharged under the visible one.

From~\eqref{toywrappings}, we see that $U(1)_{\text v}$ and $U(1)_{\text h}$ both couple to the RR axions $\phi^2$ and $\phi^3$, i.e. the reduction of $C_3$ along the cycles $[\alpha^2]$ and $[\alpha^3]$. The matrix of axionic charges under $U(1)_{\text v}\times U(1)_{\text h}$ is precisely~\eqref{charges}, hence reproducing the mass mixing mechanism described in previous sections. For generic Calabi-Yau spaces $\bf{X_6}$ the complex structure moduli space metric $G$ is under poor control, although in simple setups such as toroidal compactifications it is known (in fact, it is diagonal at tree level).

We see with this very simple model that mass mixing and the St\"uckelberg portal can be easily implemented in models with intersecting D-branes. These setups are one of the most fruitful 
frameworks
for (MS)SM-like string theory constructions. In a companion paper~\cite{us} we work out extensions of semi-realistic models by a hidden sector that communicates with the visible one through $U(1)$ mass mixing. There, we explore in detail the generic properties of such constructions and give a concrete realization in a toroidal compactification where computations can be carried out quite explicitly.

An important factor to take into account in such scenarios is that the entries of the mass matrix of $Z'$s are of order $M_{ab}\sim{\mathcal{O}}(gM_s)$, where $g$ are gauge coupling constants and $M_s$ is the string scale. In order to obtain $Z'$ bosons around the TeV range one can consider low string scale and/or anisotropic compactifications where some of the couplings $g$ are small. Here we propose a third possibility that we explore further in~\cite{us}. The presence of a large number of $U(1)$s generates a large mass matrix $M$. Upon diagonalization to the basis of physical $Z'$s, the well known eigenvalue repulsion effect may easily generate a hierarchy between the mass of the lightest $Z'$ and the string scale. A combination of these three mechanisms can lead to $Z'$ bosons in a phenomenologically interesting range.

\section{Conclusions}

In summary, we have presented a framework in which the SM sector naturally interacts with the hidden sector at the renormalizable level through mass mixings of St\"uckelberg $U(1)$ gauge fields. 
Thus, in addition to the Higgs boson, St\"uckelberg $U(1)$'s provide another unique portal into dark sectors. In contrast to other Z' mediation scenarios, our setup has added appealing features both phenomenologically and from a model building viewpoint, as
(i)  the choice of extra $U(1)$'s can be broadened 
without the need of introducing unwanted exotic matter, and
(ii) tree-level interactions between the visible and hidden sectors can be generated.
Explicit constructions of  string models exemplifying this scenario and more detailed
phenomenological studies are presented in a companion paper \cite{us}.

\subsection{Acknowledgments}

We thank Yang Bai, Lisa Everett, Jan Hajer and Ran Lu for helpful discussions.
This work 
is supported in part by the DOE grant DE-FG-02-95ER40896 and the HKRGC grant 604213. WZF is also supported by the Alexander von Humboldt Foundation.


\begin{thebibliography}{99}

\bibitem{Patt:2006fw} 
  B.~Patt and F.~Wilczek,
  hep-ph/0605188.

\bibitem{Essig:2013lka} 
  R.~Essig, J.~A.~Jaros, W.~Wester, P.~H.~Adrian, S.~Andreas, T.~Averett, O.~Baker and B.~Batell {\it et al.},
  arXiv:1311.0029 [hep-ph].

\bibitem{Langacker:2008yv} 
  P.~Langacker,
  Rev.\ Mod.\ Phys.\  {\bf 81}, 1199 (2009)
  [arXiv:0801.1345 [hep-ph]].
  
  \bibitem{Stringy-Z'}
  See, e.g., \cite{Ibanez:2012zz}. For some earlier studies on Z' bosons in heterotic string theory, see, e.g. \cite{Kakushadze:1997mc} and \cite{Cleaver:1998gc}. 
 
   
  
\bibitem{Ibanez:2012zz} 
  L.~E.~Ibanez and A.~M.~Uranga,
  Cambridge, UK: Univ. Pr. (2012) 673 p
  
\bibitem{Kakushadze:1997mc} 
  Z.~Kakushadze, G.~Shiu, S.~H.~H.~Tye and Y.~Vtorov-Karevsky,
  Int.\ J.\ Mod.\ Phys.\ A {\bf 13}, 2551 (1998)
  [hep-th/9710149].

\bibitem{Cleaver:1998gc} 
  G.~Cleaver, M.~Cvetic, J.~R.~Espinosa, L.~L.~Everett, P.~Langacker and J.~Wang,
  Phys.\ Rev.\ D {\bf 59}, 055005 (1999)
  [hep-ph/9807479].

  
   \bibitem{kinetic}
Loop-suppressed kinetic mixing also arises generically by integrating out charged massive states~\cite{Holdom:1985ag}. Its effect is subleading with respect to mass-mixing, so we neglect it in this work.

\bibitem{Holdom:1985ag} 
  B.~Holdom,
  Phys.\ Lett.\ B {\bf 166}, 196 (1986).
  
\bibitem{Blumenhagen:2005mu} 
  R.~Blumenhagen, M.~Cvetic, P.~Langacker and G.~Shiu,
  Ann.\ Rev.\ Nucl.\ Part.\ Sci.\  {\bf 55}, 71 (2005)
  [hep-th/0502005].
  
\bibitem{Blumenhagen:2006ci} 
  R.~Blumenhagen, B.~Kors, D.~Lust and S.~Stieberger,
  Phys.\ Rept.\  {\bf 445}, 1 (2007)
  [hep-th/0610327].
  
\bibitem{Marchesano:2007de} 
  F.~Marchesano,
  Fortsch.\ Phys.\  {\bf 55}, 491 (2007)
  [hep-th/0702094 [HEP-TH]].
  
    
     
  

\bibitem{us} 
  W.~-Z.~Feng, G.~Shiu, P.~Soler and F.~Ye,
  JHEP {\bf 1405}, 065 (2014)
  [arXiv:1401.5890 [hep-ph]].
  
  


\bibitem{Banks:2010zn} 
  T.~Banks and N.~Seiberg,
  Phys.\ Rev.\ D {\bf 83}, 084019 (2011)
  [arXiv:1011.5120 [hep-th]].
 

  
\bibitem{Shiu:2013wxa} 
  G.~Shiu, P.~Soler and F.~Ye,
  Phys.\  Rev.\  Lett.\  {\bf 110}, 241304 (2013)
  [arXiv:1302.5471 [hep-th]].

 \bibitem{Anomaly}
We require no mixed $U(1)_{\text{v}}$-gravitational anomalies, and furthermore no $U(1)_Y-U(1)^2_{\text{v}}$ anomalies, which can always be eliminated by combining $U(1)_{\text{v}}$ with hypercharge~\cite{Ibanez:1999it}.
 
\bibitem{Ibanez:1999it} 
  L.~E.~Ibanez and F.~Quevedo,
  JHEP {\bf 9910}, 001 (1999)
  [hep-ph/9908305].
 
\bibitem{Alves:2013tqa} 
  A.~Alves, S.~Profumo and F.~S.~Queiroz,
  JHEP {\bf 1404}, 063 (2014)
  [arXiv:1312.5281 [hep-ph]].
    
\bibitem{Arcadi:2013qia} 
  G.~Arcadi, Y.~Mambrini, M.~H.~G.~Tytgat and B.~Zaldivar,
  JHEP {\bf 1403}, 134 (2014)
  [arXiv:1401.0221 [hep-ph]].
  

 \bibitem{anomalous}
   Notice that anomalous $U(1)$'s yield some particular couplings that may lead to specific experimental signatures \cite{Anastasopoulos:2006cz,Kumar:2007zza}.
  
\bibitem{Anastasopoulos:2006cz} 
  P.~Anastasopoulos, M.~Bianchi, E.~Dudas and E.~Kiritsis,
  JHEP {\bf 0611}, 057 (2006)
  [hep-th/0605225].
  
\bibitem{Kumar:2007zza} 
  J.~Kumar, A.~Rajaraman and J.~D.~Wells,
  Phys.\ Rev.\ D {\bf 77}, 066011 (2008)
  [arXiv:0707.3488 [hep-ph]].


\bibitem{Ghilencea:2002da} 
  D.~M.~Ghilencea, L.~E.~Ibanez, N.~Irges and F.~Quevedo,
  JHEP {\bf 0208}, 016 (2002)
  [hep-ph/0205083].
 
 
 
 \bibitem{invariants}
 Another way to state this is that terms in the Lagrangian must be gauge invariant under the original symmetries $U(1)_{\text{v}}$ and $U(1)_{\text{h}}$, and hence can only include singlets from each sector.



\bibitem{Langacker:2007ac} 
  P.~Langacker, G.~Paz, L.~-T.~Wang and I.~Yavin,
  Phys.\ Rev.\ Lett.\  {\bf 100}, 041802 (2008)
  [arXiv:0710.1632 [hep-ph]].
  
\bibitem{Langacker:2008ip} 
  P.~Langacker, G.~Paz, L.~-T.~Wang and I.~Yavin,
  Phys.\ Rev.\ D {\bf 77}, 085033 (2008)
  [arXiv:0801.3693 [hep-ph]].
  
\bibitem{Verlinde:2007qk} 
  H.~Verlinde, L.~-T.~Wang, M.~Wijnholt and I.~Yavin,
  JHEP {\bf 0802}, 082 (2008)
  [arXiv:0711.3214 [hep-th]].
  
  
  \bibitem{Nath}
 Mass mixing between hypercharge and a hidden $U(1)$ as studied in~\cite{Kors:2004dx,Cheung:2007ut,Feldman:2007wj} is highly constrained from the appearance of fractional charges in the hidden sector~\cite{Shiu:2013wxa,Langacker:2011db}.

 
\bibitem{Kors:2004dx} 
  B.~Kors and P.~Nath,
  Phys.\ Lett.\ B {\bf 586}, 366 (2004)
  [hep-ph/0402047].
  
  
\bibitem{Cheung:2007ut} 
  K.~Cheung and T.~-C.~Yuan,
  JHEP {\bf 0703}, 120 (2007).
  
\bibitem{Feldman:2007wj} 
  D.~Feldman, Z.~Liu and P.~Nath,
  Phys.\ Rev.\ D {\bf 75}, 115001 (2007)
  [hep-ph/0702123 [HEP-PH]].
  

\bibitem{Langacker:2011db} 
  See, e.g. P.~Langacker and G.~Steigman,
  Phys.\ Rev.\ D {\bf 84}, 065040 (2011)
  [arXiv:1107.3131 [hep-ph]],  and references therein.


  
  
  
  
  
 
\bibitem{Strassler:2006im} 
  M.~J.~Strassler and K.~M.~Zurek,
  Phys.\ Lett.\ B {\bf 651}, 374 (2007)
  [hep-ph/0604261].

\bibitem{Strassler:2006qa} 
  M.~J.~Strassler,
  hep-ph/0607160.
  
\bibitem{Strassler:2006ri} 
  M.~J.~Strassler and K.~M.~Zurek,
  Phys.\ Lett.\ B {\bf 661}, 263 (2008)
  [hep-ph/0605193].
  
\bibitem{Han:2007ae} 
  T.~Han, Z.~Si, K.~M.~Zurek and M.~J.~Strassler,
  JHEP {\bf 0807}, 008 (2008)
  [arXiv:0712.2041 [hep-ph]].
    





 
 

 
 
 
  

  

  
\bibitem{Jaeckel:2010ni} 
  J.~Jaeckel and A.~Ringwald,
  Ann.\ Rev.\ Nucl.\ Part.\ Sci.\  {\bf 60}, 405 (2010)
  [arXiv:1002.0329 [hep-ph]].
  
  
\end{thebibliography}
\end{document}